\documentclass[3p,twocolumn]{elsarticle}

\usepackage{amssymb}
\usepackage{amsmath}
\usepackage{lineno}
\journal{Chaos Solitons and Fractals}

\begin{document}


\begin{frontmatter}

\title{Strategy dependent learning activity in cyclic dominant systems}

\author[label1]{Attila Szolnoki}
\author[label2]{Xiaojie Chen}

\address[label1]{Institute of Technical Physics and Materials Science, Centre for Energy Research, Hungarian Academy of Sciences, P.O. Box 49, H-1525 Budapest, Hungary}

\address[label2]{School of Mathematical Sciences, University of Electronic Science and Technology of China, Chengdu 611731, China}

\begin{abstract}
The prototype of a cyclic dominant system is the so-called rock-scissors-paper game, but similar relation among competing strategies can be identified in several other models of evolutionary game theory. In this work we assume that a specific strategy from the available set is reluctant to adopt alternative states, hence the related learning activity is reduced no matter which other strategy is considered for adoption. Paradoxically, this modification of the basic model will primarily elevate the stationary fraction of another strategy who is the virtual predator of the one with reduced learning activity. This general reaction of the studied systems is in agreement with our understanding about Lotka-Volterra type cyclic dominant systems where lowering the invasion rate between a source and target species promotes the growth of former population. The observed effect is highly non-linear because the effective invasion rates between strategies may depend sensitively on the details of the actual model.
\end{abstract}


\end{frontmatter}

\section{Introduction}
\label{intro}

Cyclic dominant relation among competing species is a well-known and frequently quoted mechanism when we want to understand the diversity of species in a living system which evolves under the rule of Darwinian selection principle \cite{szolnoki_jrsif14,frey_pa10,mobilia_jtb10,park_c18,nagatani_srep18,avelino_epl18,lutz_g17,amaral_prsa20}. In the simplest and highly simplified case a species $A$ dominates species $B$, but he latter outperforms species $C$. More interestingly, species $C$ invades species $A$, which establishes a fine balance among all participants. This relation is in the heart of the celebrated rock-scissors-paper ($RSP$) game, which was identified in many biological and social systems, ranging from interaction between bacterias \cite{paquin_n83,kerr_n02,kosakovski_pone18}, plants \cite{cameron_jecol09}, animals \cite{buss_an79,sinervo_n96,guill_jtb11}, and even humans \cite{mobilia_g16,ed_19,szolnoki_csf20,nagatani_jtb19,szolnoki_pre17,li_k_epl20}.

A seemingly surprisingly consequence of the non-transitive dynamics is when the invasion rates between species are unequal. Here a lowered invasion rate by one species results in a decreased stationary fraction of the species that invades it. As a consequence, due to odd number of species in the loop, by lowering the invasion rate of a species will increase its fraction in the population \cite{frean_prsb01,avelino_pre19b}.

Before we proceed, we should note that the mentioned non-transitive relation between competing partners may emerge in other types of evolutionary game models, too \cite{szolnoki_pre10b,canova_jsp18,cheng_f_amc20,li_qr_pa20,liu_xs_epl19,rong_zh_c19,takeshue_epl19,quan_j_c19,shao_yx_epl19}. For instance, by introducing voluntary participation into the classic public goods game we can observe an $RSP$-type cycle among cooperators, defectors, and loners \cite{hauert_s02}. But similar phenomenon can also be identified in the simplest prisoner's dilemma game \cite{szabo_pre02b}. There are other examples of more complex models as well where the relation of strategies is less obvious, but the collective behavior of spatial setting results in similar propagating fronts patterns we have already observed for $RSP$ game. Beside the obvious parallel to the latter, however, there is a significant difference in the lastly mentioned evolutionary game models. In particular, staying at the simplest prisoner's dilemma extended by loner strategy, a cooperator agent may adopt not just defector, but also loner strategy during an elementary step. In this way there is no a one-way flow of invasion among strategies that we experienced for $RSP$-game. 

In the following we explore the possible consequence of modified invasion rates in these models where there is bidirectional adoption between competing strategies. Technically it can be done in the simplest way if we introduce a strategy-dependent learning activity. More precisely, we assume that a player who posses one of the strategies are reluctant to change their states and the usual payoff-difference driven adoption probability is multiplied by a constant factor to them. Our primary interest is to check how this intervention into the dynamical rule changes the stationary fractions of strategies.

In the next section we define the general scheme that is applied to specific models. After we study in detail the consequences for three previously suggested models where non-transitive relations were identified. Last we conclude with the summary of the results and a discussion of their implications.

\section{Strategy-dependent learning activity in cyclically dominant systems}
\label{def}

\begin{figure}[h!]
\centering
\includegraphics[width=4.5cm]{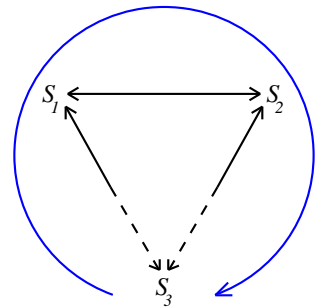}\\
\caption{General scheme of the model modification. In the basic model three strategies compete where there is an effective cyclic dominance between them, as signed by a blue arrow. In particular $S_1$ invades $S_2$, who invades $S_3$, while the latter invades $S_1$. In the modified model $S_3$ will adopt an external strategy with a reduced probability. This is marked by dashed arrows in the plot.}\label{general}
\end{figure}

According to the general setup, we have at least three competing strategies who invade each other in a cyclic manner. This non-transitive relation between $S_1, S_2$, and $S_3$ strategies is marked by a blue arrow in Fig.~\ref{general}. According to the broadly accepted dynamical rule a player $y$ having strategy $s_y$ adopts the strategy of a neighboring player $x$ who has $s_x$ strategy with a probability driven by the payoff difference of the mentioned players:
\begin{equation}
\Gamma (s_x \to s_y) = [1+\exp(\Pi_{s_y}-\Pi_{s_x})/K]^{-1}\,.
\end{equation}
Here $\Pi_{s_x}$ and $\Pi_{s_y}$ are the payoff values of related players who collected it by playing the actual game with their neighbors. Furthermore $K$ noise parameter quantifies a certain level of uncertainty in strategy adoptions in a way that it is likely to adopt the strategy of a player who has higher payoff value, but the reversed process may also happen with a low probability.

As we have already stressed in the introduction, strategy invasion may happen in both directions between competing strategies that is a fundamental difference from the classic $RSP$ game. Furthermore, we assume that players, having a specific strategy, are reluctant to change their state. In the mentioned plot $S_3$ denotes this specific strategy. Accordingly, a player having $S_3$ strategy adopts an alternative strategy with a probability $w_l \cdot \Gamma$, where $\Gamma$ is the payoff difference driven adoption rate while $0 < w_l \le 1$ represents the learning activity of players following strategy $S_3$. This intervention into the microscopic rule is marked by dashed arrows in Fig.~\ref{general}.

For easier reference in the relation of $S_2$ and $S_3$ we will call $S_2$ as a ``predator" and $S_3$ as a ``prey" strategy because in most of the cases $S_2$ invades $S_3$. But we highlight that the direction of this invasion is not exclusive. Furthermore we note that the reduced learning activity players with $S_3$ strategy is also valid between $S_1$ and $S_3$ strategy invasion that is marked by dashes arrows in the mentioned plot.

In the following we revisit three previously studied models where an effective cyclic dominance was revealed among the strategies who evolve according to an evolutionary game theory rule. In every case we briefly summarize the relation of competing strategies and explore the possible consequence of modified dynamical rule on the stationary fractions of competitors. Notably, the game is played on a square grid, but the results remain intact if we apply alternative interaction graphs. For proper comparison with previous results we apply $K=0.1$ value for all discussed cases, but naturally our observations are not restricted to this noise level. Further details of Monte Carlo simulations including system sizes, relaxation and measuring time will be specifically provided to each case. 

\subsection{Case study 1: Prisoner's dilemma with loners}

We first consider the traditional prisoner's dilemma game where beside cooperator ($C$) and defector ($D$) strategies loners ($L$) are also involved \cite{szabo_pre02b}. The payoff of the mentioned strategies is

\begin{figure*}
\centering
\includegraphics[width=3.5cm]{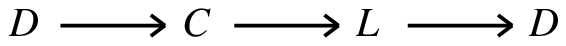}\\
\includegraphics[width=5.5cm]{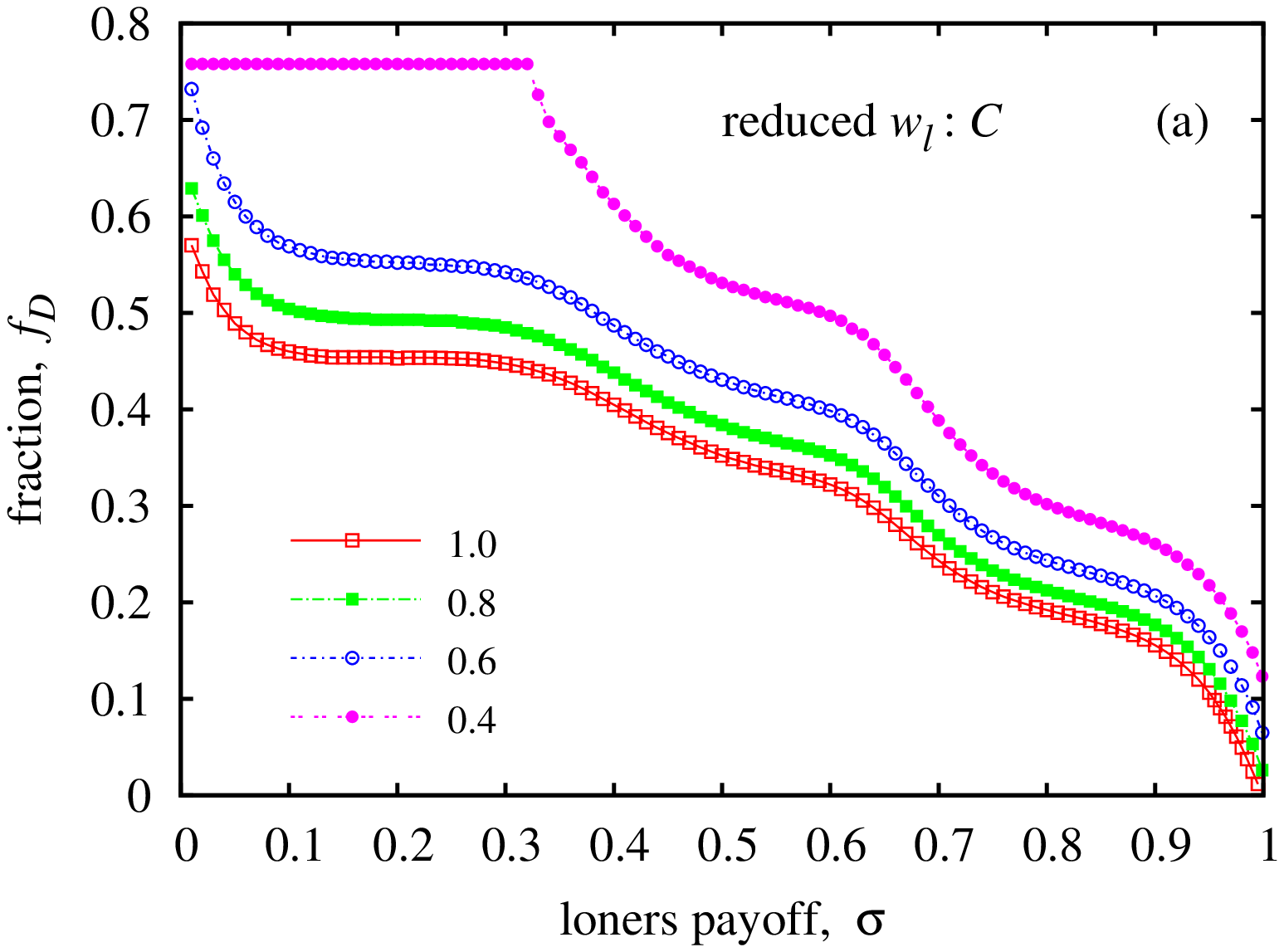}\includegraphics[width=5.5cm]{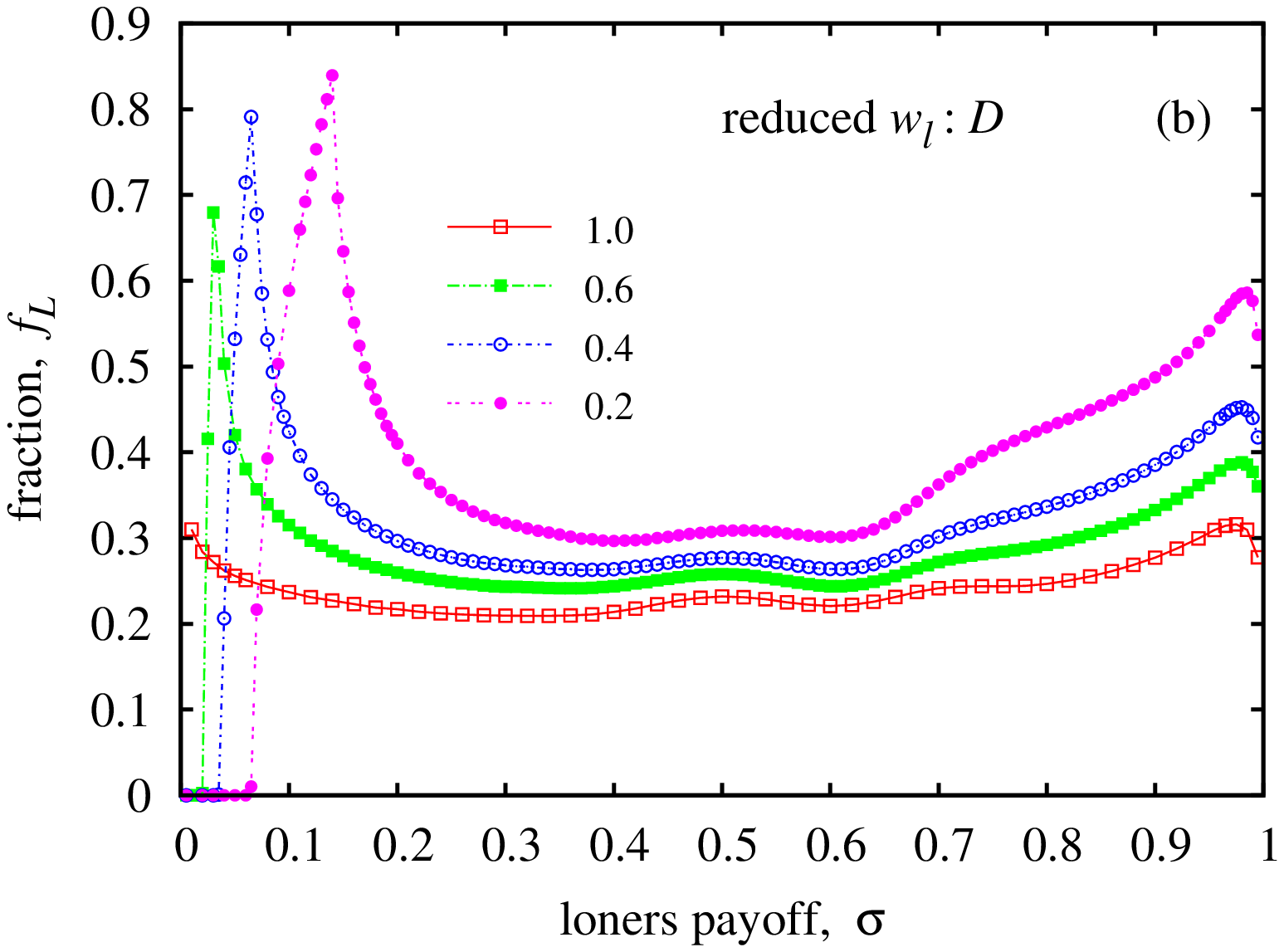}\includegraphics[width=5.5cm]{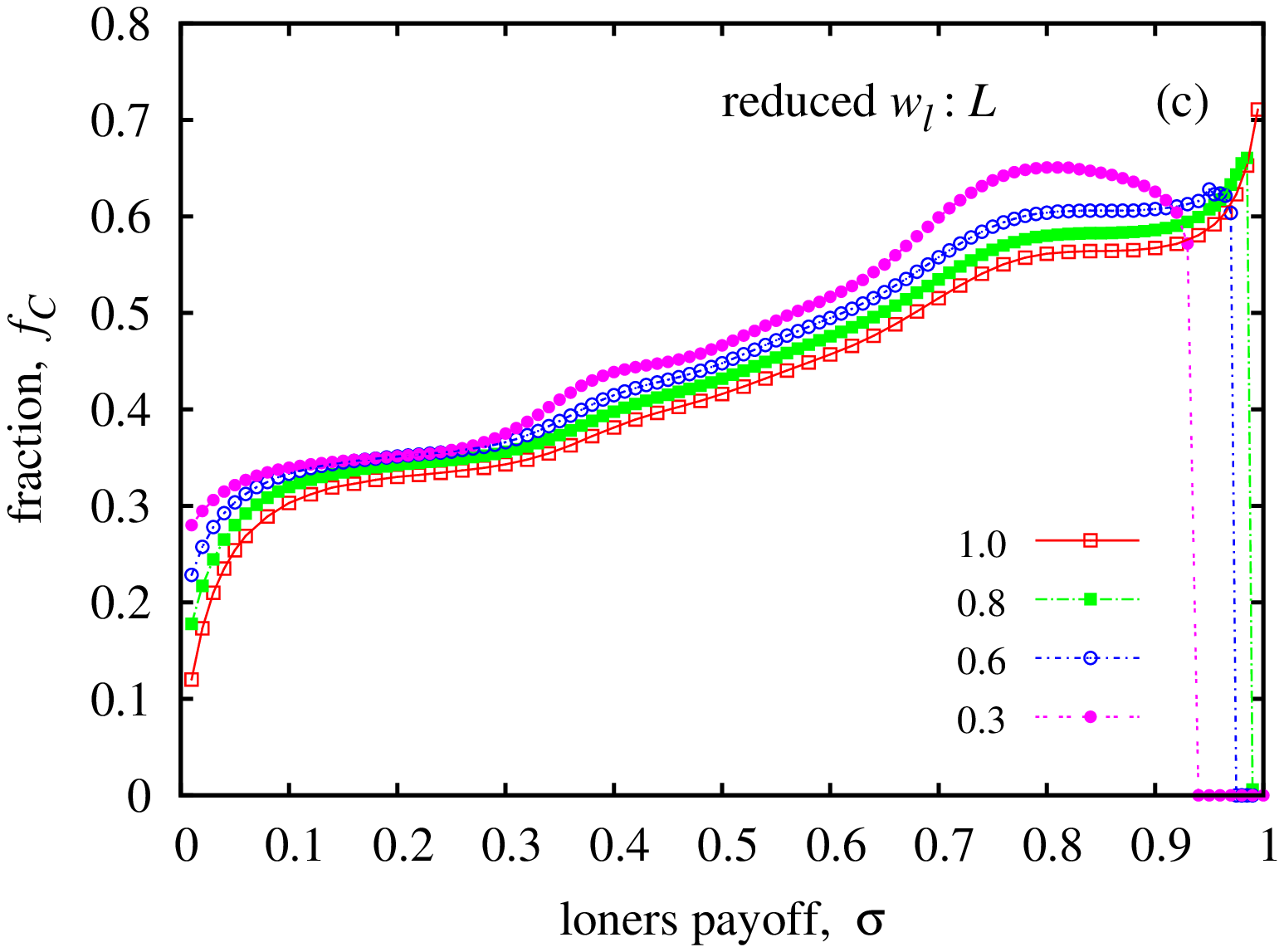}\\
\caption{Improvement of ``predator" strategy in dependence of loner's payoff at $T=1.1$ in prisoner's dilemma game where cooperators and defectors compete with loners. The applied $w_l$ learning activities for a specified strategy are marked in the figure legends. For example in panel~(a) cooperators are forced to adopt external strategies with lower probabilities. As a response, the fraction of defectors increases as we decrease the value of $w_l$. The upper plot marks the ranks of strategies, where defectors beat cooperators who beat loners, etc. We note that the improvement of predator strategy is only valid if the three-strategy state is maintained. At small $\sigma$ in panel~(b) or at large $\sigma$ in panel~(c) the system leaves the cyclically dominant solution.}\label{loner}
\end{figure*}
\vspace{0.2cm}

\centerline{\begin{tabular}{r|c c c}
{\bf } & $D$ & $C$ & $L$\\
\hline
$D$ & 0 & T & $\sigma$\\
$C$ & 0 & 1 & $\sigma$\\
$L$ & $\sigma$ & $\sigma$ & $\sigma$\\
\end{tabular}
}

In this model beside the temptation to defect ($T$) the key parameter is $\sigma$ that determines a loner's and its partner's payoff. According to previous observations the presence of risk averse loners introduce a cyclic dominance among competitors which help the system to avoid the expected tragedy of the commons state. More precisely, while defectors do better than cooperators, but they are unable to exploit loners. On the other hand, cooperators can help each other better than loners, which establishes the mentioned non-transitive relation.

When we apply a reduced learning activity to players who carry a specific strategy then the stationary faction of competing strategies change. In Figure~\ref{loner} we have summarized the most important effects where each panel shows the consequence when a certain strategy is ``protected" by a lowered invasion rate. For example in panel~(a) cooperators are reluctant to change, still the most rewarded players are defectors, who are the ``predators" of $C$ strategy in the cyclic loop. The figure shows the stationary fraction in dependence of loner's payoff for different strategy learning activity as indicated in the legend, where the largest change can be observed for small $w_l$ values. Notably, neither $f_C$ nor $f_L$ growth in response to lower $w_l$ values of cooperators. Actually, the opposite is true, the fractions of the mentioned strategies decrease due to the conservation law of total strategy number.

Similar behavior can be seen in panel~(b) where defectors are reluctant to change, but loners enjoy this intervention into the model dynamics. Here, at low $\sigma$ values we can see that the described increment of $f_L$ is not valid anymore, but this is because the system leaves the three-strategy state at this parameter value. Consequently the argument, which is based on the cyclic loop of ranks, cannot be applied anymore. In the last panel loners adopts external strategy with a lowered probability that elevates the general cooperation level. But this effect vanishes at large $\sigma$ values when the cyclically dominant solution is not valid anymore.

For the requested accuracy of data we have used $L \times L = 800 \times 800$ system size where $4 \cdot 10^5$ Monte Carlo steps (MCS) were applied to reach the stationary state at a given parameter value and we averaged stationary data over another $6 \cdot 10^5$ $MCS$s.

\begin{figure}
\centering
\includegraphics[width=7.5cm]{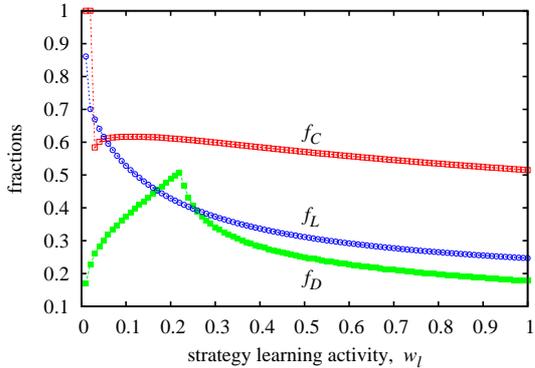}\\
\caption{Stationary fraction of ``predator" strategies in dependence of the learning activity of their related ``prey" strategies. It is a general phenomenon that a lower $w_l$ of ``prey" strategy results in a higher stationary fraction of its ``predator" strategy if the system remains in the three-strategy cyclically dominant solution. If it is not fulfilled then a two-strategy solution is still possible, as it is illustrated for $f_D$ curve at small $w_l$ values. The cost of loners is $\sigma=0.7$ for $f_C$, 0.8 for $f_L$, and 0.85 for $f_D$ curve while $T=1.1$ for all cases.}\label{loner_w}
\end{figure}

In Fig.~\ref{loner_w} we have plotted the mentioned ``predator" strategy fractions for the three different cases as a function of $w_l$ value. All curves indicate clearly that the portion of ``predator" strategy increases continuously by lowering the $w_l$ of ``prey" strategy. This observation is valid until the three-strategy solution remains stable. The latter is not fulfilled at low $w_l$ values for $f_D$ curve where loners die out and cooperators and defectors coexist. Notably the applied $T=1.1$ value would be too high for their coexistence in the traditional model, but the lowered $w_l$ value shifts the balance for the benefit of cooperator strategy.

Summing up, in the simplest evolutionary game theory model where the applied payoff values indicate a cyclic loop among competing strategies the introduction of reduced strategy learning activity results in a similar system reaction we previously observed for the traditional $RSP$ model where invasion was only possible in a directed way.

\subsection{Case study 2: Conditional cooperators and defectors with deceitful defectors}

In our next example we stay at prisoner's dilemma game but introduce more sophisticated strategies \cite{szolnoki_njp14}. Here, beside pure defectors ($D$) we have conditional cooperators ($C$) who successfully detect defection with probability $p$ and avoid being exploited. As a tactical response, we allow some defectors to hide their proper face and deceive cooperator players. They are called deceitful defectors ($X$) who can still exploit cooperator mates. Of course there is no free lunch, hence deceitful strategy should bear an extra cost of successful hide. The applied payoff matrix is the following:

\vspace{0.2cm}

\begin{tabular}{r|c c c}
{\bf } & $D$ & $C$ & $X$\\
\hline
$D$ & 0 & 0 & 0\\
$C$ & 0/S & 1 & $S$\\
$X$ & $-\gamma$ & $(T-\gamma)$ & $-\gamma$\\
\end{tabular}

\vspace{0.2cm}

where $\gamma$ denotes the mentioned extra cost for strategy $X$, while $T$ (temptation to defect) and $S$ (sucker's payoff) are the usual free parameters of the traditional prisoner's dilemma game. For the proper comparison with previous results, but without loosing generality, we apply $T=1.3$ and $S=-0.3$ in the following. Importantly, a cooperator successfully recognizes pure defectors with probability $p$, hence in this case the zero value is used in the second row, first column when a $C$ player meets with a $D$ player. With probability $1-p$, however, a $C$ player fails to uncover bad intention, hence the usual $S$ value should be considered.

A previous work revealed that the mentioned strategies are capable to form a cyclically dominant loop if the detecting probability $p$ is not too small and the cost of hiding $\gamma$ is not too large. \cite{szolnoki_njp14}. An interesting observation was that deceitful behavior fares better if it is costly, which was an excellence consequence of cyclically dominant ranks of the mentioned strategies. In particular, the increased $\gamma$ value would be believed to weaken strategy $X$, still the biggest looser is its ``predator" strategy, which is $D$. Indeed, in this model at reasonably high $p$ value conditional cooperators outperform pure defectors who dominate deceitful defectors who should bear the extra cost. But the latter pays when an $X$ player meets with a $C$ player. 

In our present work we study how the mentioned cyclic loop influences the stationary fractions of strategies when one of them is technically  ``strengthened" by a smaller transition rate. According to the na{\" \i}ve expectation a reduced learning activity helps these players to keep their strategies longer hence a time average of their fractions will grow. As we will demonstrate, however, this is not the case and the observed behavior is strongly related to the non-transitive ranks of competitor strategies. 

But before we proceed we clarify the technical details of simulations. Again, for proper comparison we applied square grid interaction topology with $L \times L = 3000 \times 3000$ system size, where typically $5 \cdot 10^5$ $MCS$s were applied for relaxation, but conceptually similar behavior can be also observed in other graphs. Notably, the unusual large system size was necessary to reach the proper cyclically dominant solution from a random initial condition at every parameter values.   

\begin{figure*}
\centering
\includegraphics[width=3.5cm]{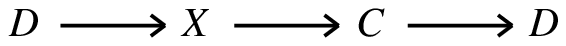}\\
\includegraphics[width=5.5cm]{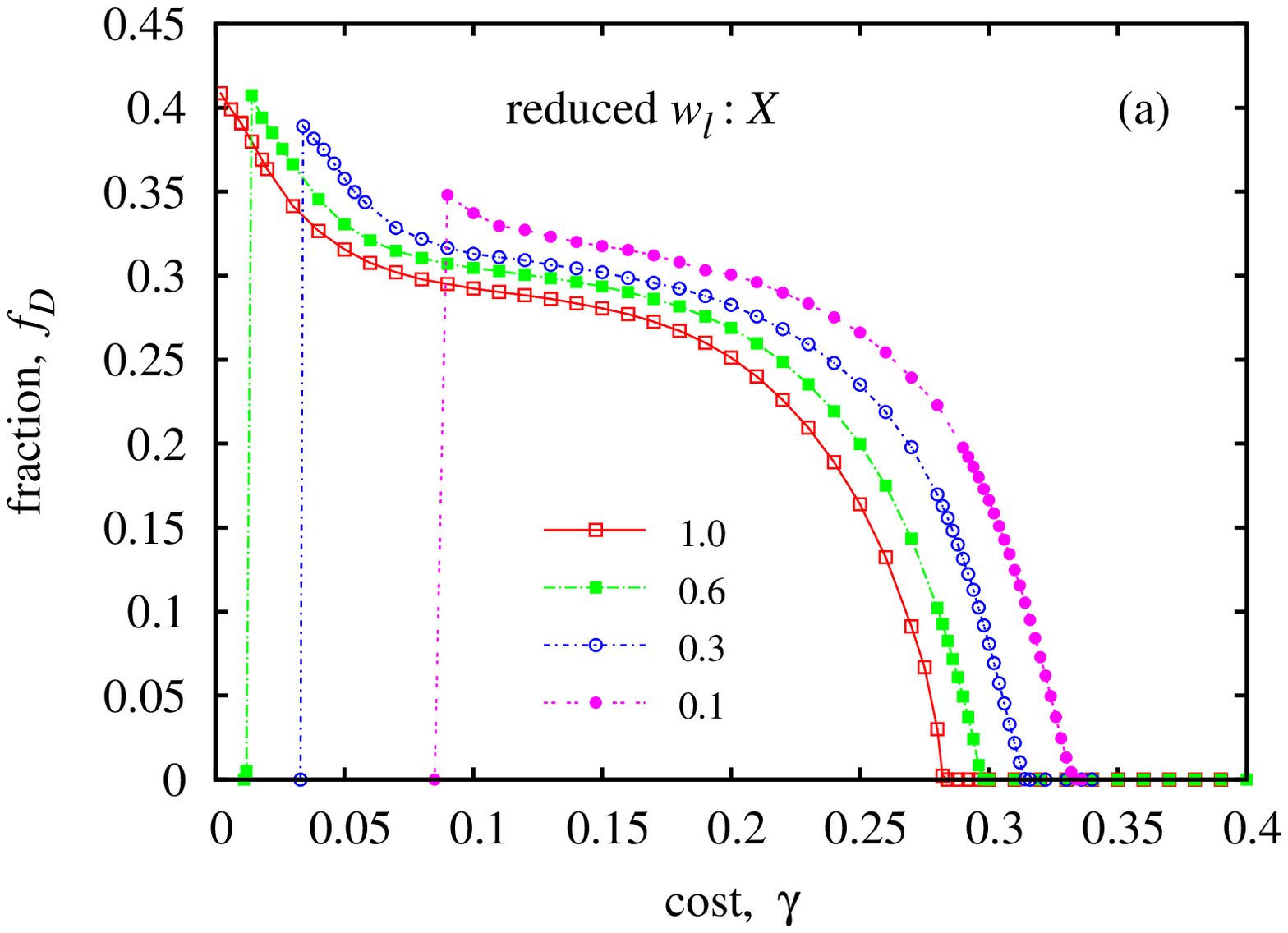}\includegraphics[width=5.5cm]{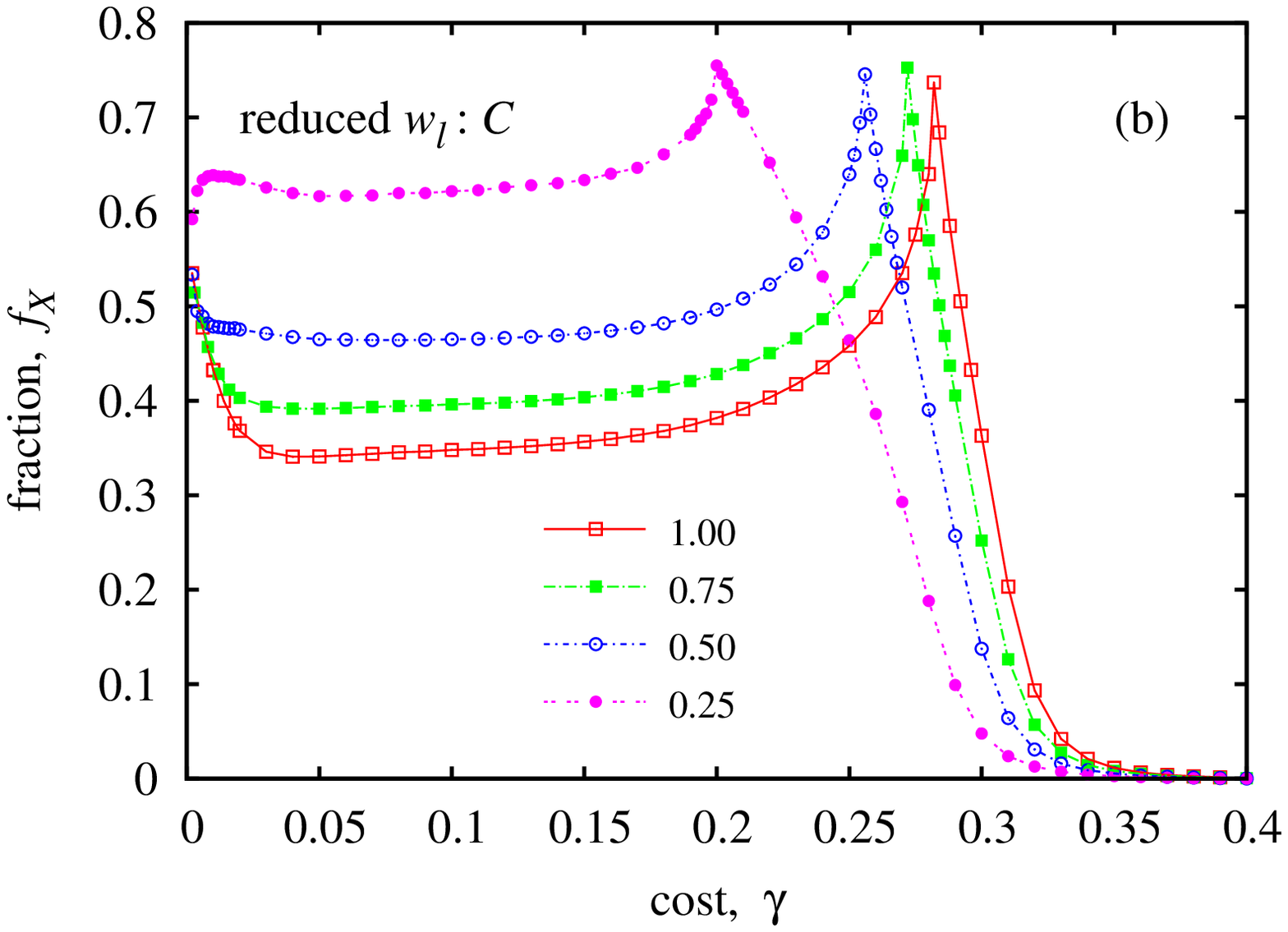}\includegraphics[width=5.5cm]{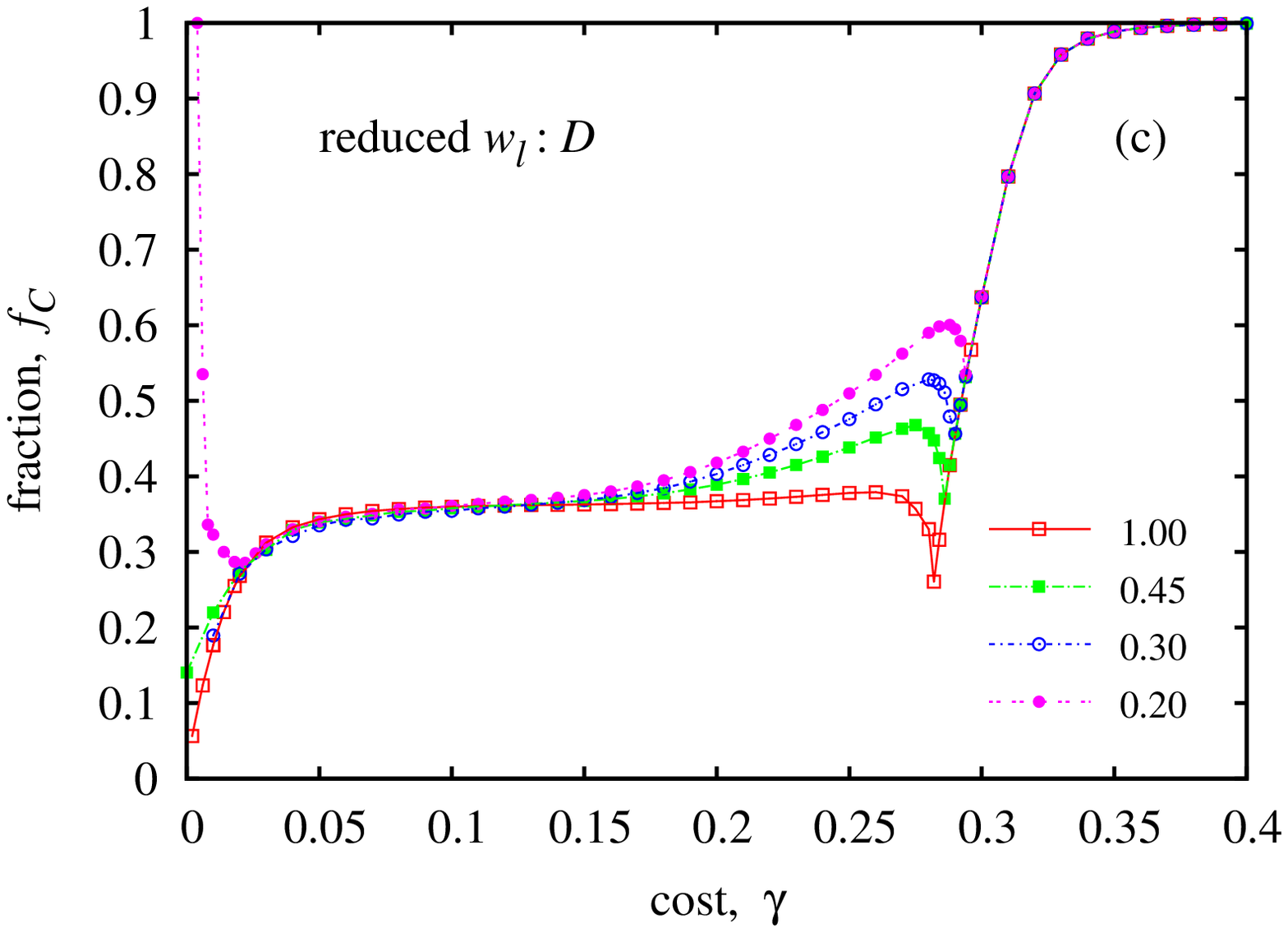}\\
\caption{Improvement of ``predator" strategy as a function of cost of deceitful defectors. In this model pure defectors ($D$), conditional cooperators ($C$), and deceitful defectors ($X$) compete. The latter has to bear an extra cost $\gamma$ to hide its proper strategy. The ranks of strategies are indicated on the top of the plot. Similarly to the previous model the ``predator" strategy enjoys the reduced learning activity of related ``prey" strategy. The latter strategy is marked in every panel. The enhanced ``predator" abundance remains valid until the system remains in the three-strategy state, where cyclic dominance is maintained. Parameters are $T=1.3, S=-0.3$, and $p=0.8$ where the latter is the probability that a conditional cooperator correctly identifies pure defectors and avoid being exploited by them.}\label{hide}
\end{figure*}

Our key observations are summarized in Fig.~\ref{hide}, where we plotted how the frequency of the specific ``predator" strategy changes in dependence of hiding cost $\gamma$ when the related ``prey" strategy uses a reduced learning activity. As for the previous model, on the top of this figure we have displayed the proper ranks of strategies. Based on this it becomes clear why a reduced learning activity supports primarily the related ``predator" strategy even if this decreased invasion rate is applied for both directions of other external strategies. This symmetry, however, which was highlighted by dashed arrows in Fig.~\ref{general}, becomes marginal in the presence of cyclic loop. The latter is proved to be so strong circumstance that we practically observe similar system reaction to those were previously reported for uni-directional invasions in the framework of $RSP$ game.

In particular, as it is shown in panel~(a), when deceitful defectors are supported by keeping their state via reduced $w_l$, then the fraction of pure defectors will increase significantly. Similarly, panel~(b) illustrates that the unambiguous victor of reduced learning activity of cooperators is deceitful defectors. This effect becomes invalid only for high $\gamma$ cost values where the system leaves the cyclically dominant three-strategy solution and only conditional cooperators coexist with deceitful defectors. Last, panel~(c) shows the system reaction when defectors adopts other strategies with reduced probability. Here the increment of conditional cooperators seem to be modest, but our goal was to present results obtained at identical $p$ value for all panels. Indeed, higher growth of $f_C$ can be observed at other parameter values.

\begin{figure}[h!]
\centering
\includegraphics[width=7.5cm]{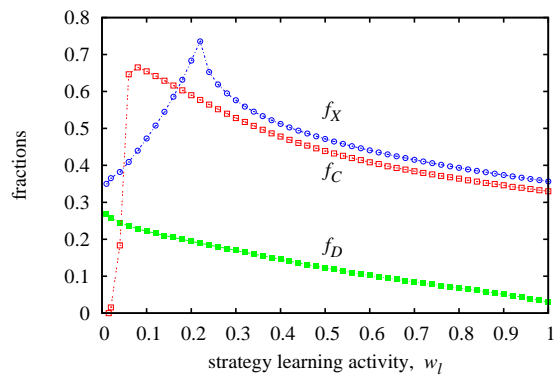}\\
\caption{Stationary fraction of ``predator" strategies in dependence of the learning activity of their related ``prey" strategies in the conditional cooperator -- deceitful defector model. Similarly to the previous model lower $w_l$ of ``prey" strategy provides a higher stationary fraction for its ``predator" strategy until the three-strategy cyclically dominant solution remains stable. The extra cost of deceitful defectors is $\gamma=0.28$ for $f_C$, and for $f_D$, while $\gamma=0.15$ for $f_X$. Other parameters are $T=1.3, S=-0.3$, and $p=0.8$.}\label{hide_w}
\end{figure}

Our next plot, shown in Fig.~\ref{hide_w}, illustrates how the frequency of ``predator" strategy changes as we decrease the learning activity of ``prey" strategy. This behavior is in agreement with the one we previously observed for the other model. More precisely, the more we ``support" a prey strategy to keep its state, the better is for the related ``predator" partner. Hence lower $w_l$ is beneficial for involved ``predators" except the small $w_l$ interval where the system evolves to a non-cyclical solution. The message, however, is clear: even if we apply more sophisticated strategies in the original prisoner's dilemma, if a closed loop of ranks emerges then we can observe system reaction similar to the simplest $RSP$ game.

\subsection{Case study 3: Multi-cycles with informed strategies}

\begin{figure*}
\centering
\includegraphics[width=3.5cm]{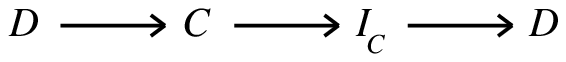}\\
\includegraphics[width=5.5cm]{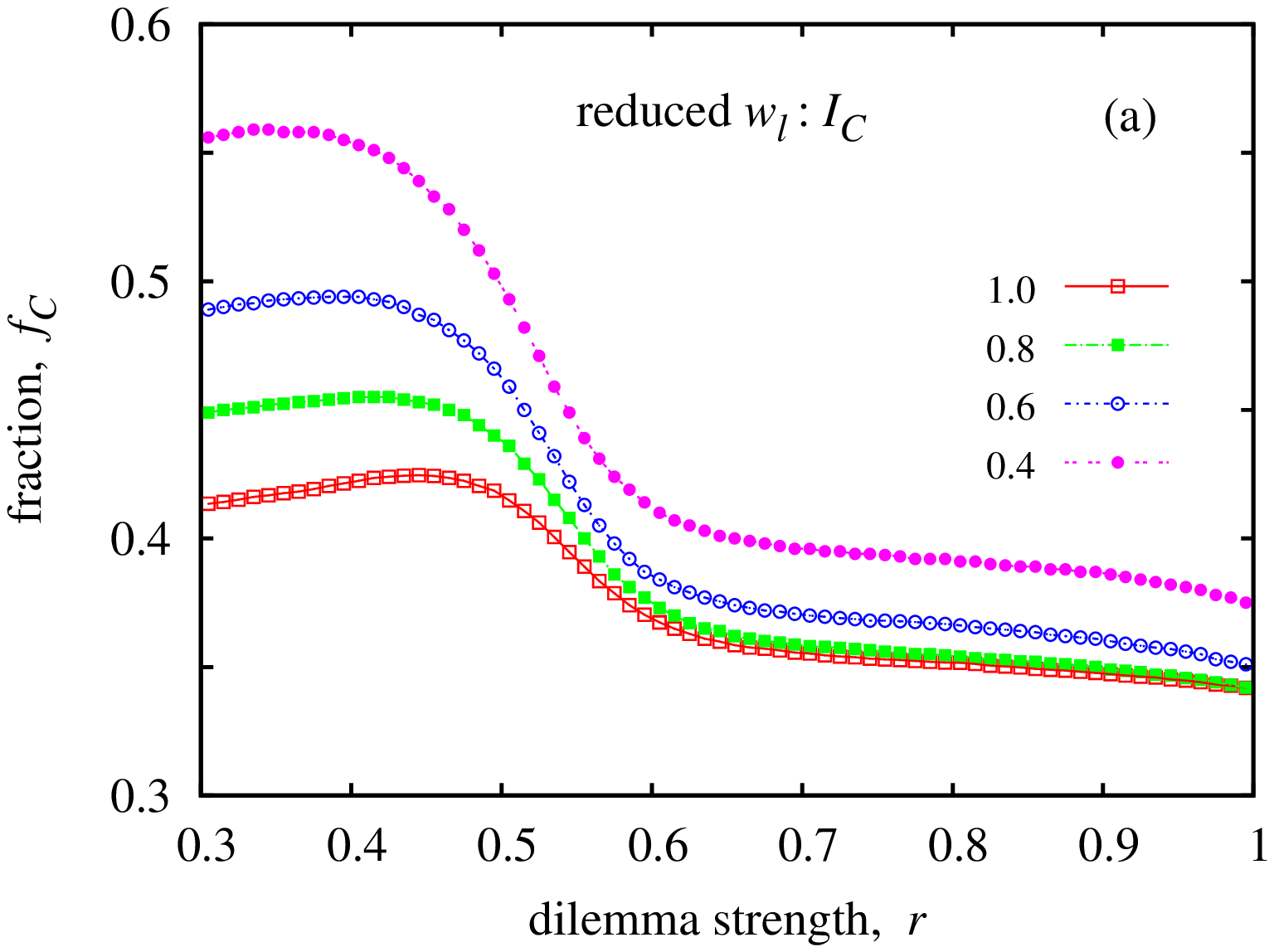}\includegraphics[width=5.5cm]{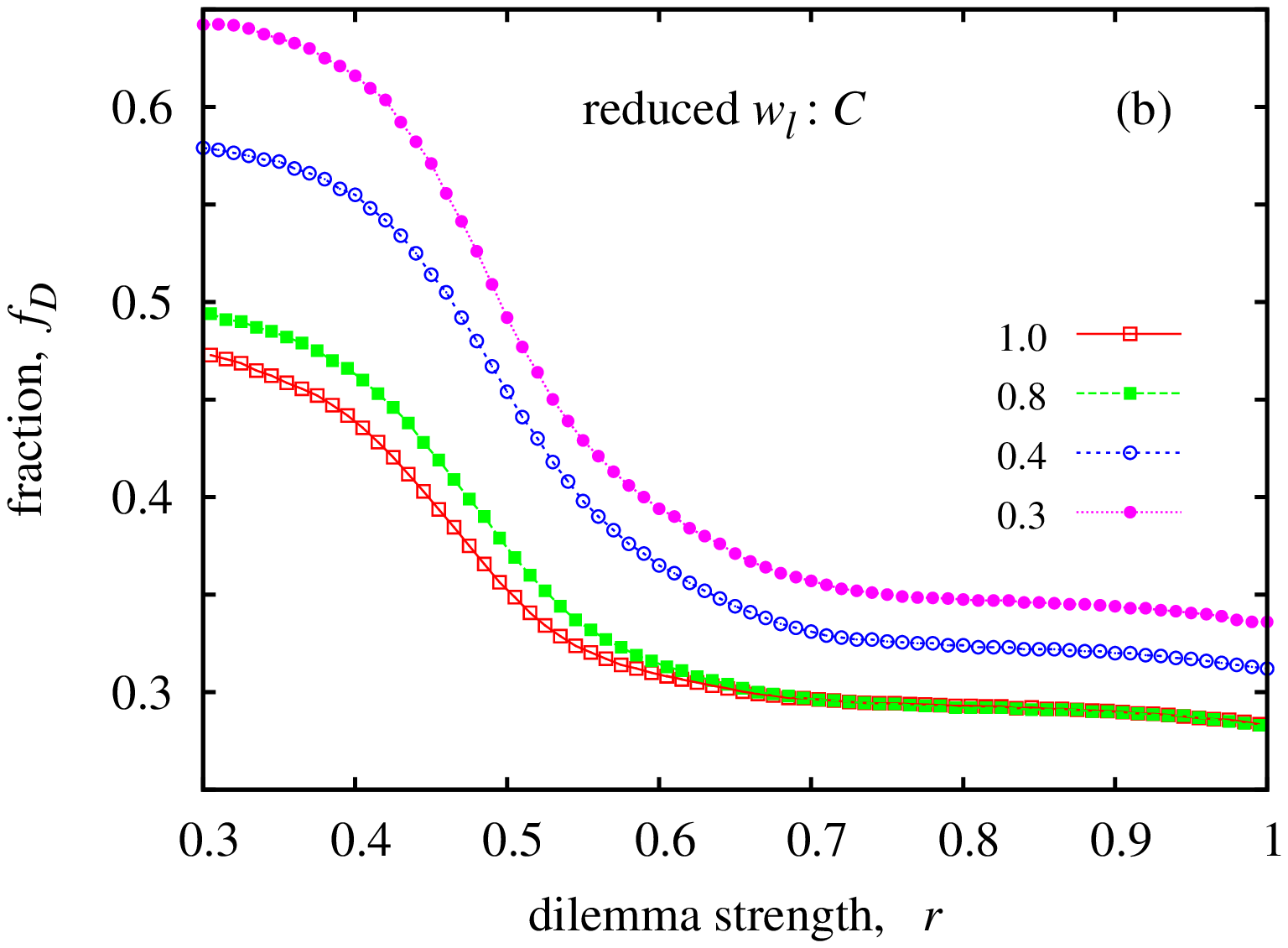}\includegraphics[width=5.5cm]{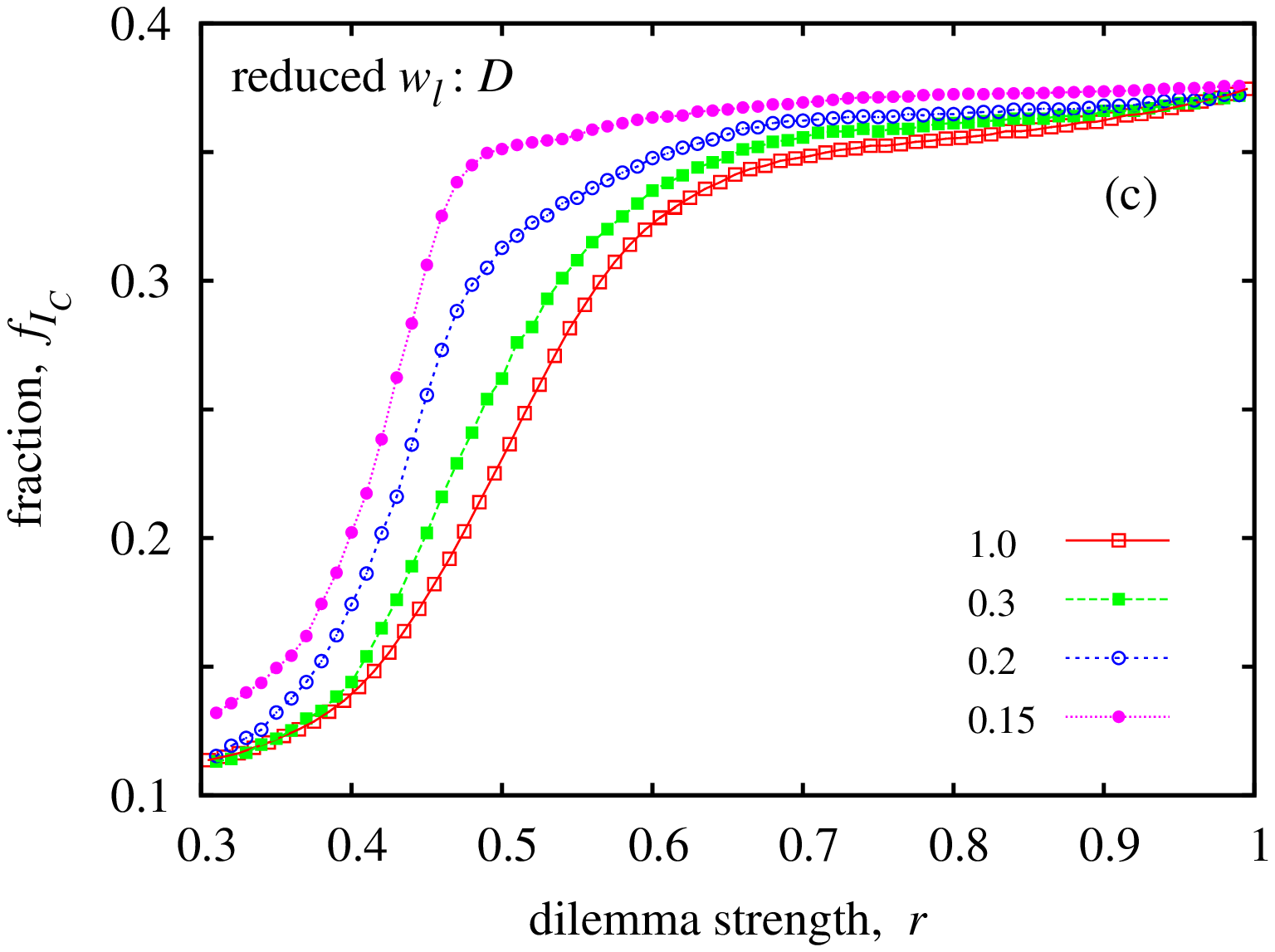}\\
\caption{Improvement of ``predator" strategy as a function of dilemma strength in the model where pure cooperator and defector strategies compete with informed cooperators. The ranks of strategies are indicated on the top of the plot, while the applied $w_l$ values are denoted in the legend. All panels illustrate clearly that the impact of reduced learning activity on the abundance of ``predator" strategy is a general phenomenon.
The value of punishment is $P=-0.3$, while the additional cost of informed cooperators is $\epsilon=0.3$ for all cases.}\label{inf_I}
\end{figure*}

Our last example is related to a system where more than three strategies are available, and not only one, but two independent solutions possess cyclic loop \cite{szolnoki_epl15}. Here, beside unconditional cooperators ($C$) and unconditional defectors ($D$) we have the so-called informed strategies who make special effort to explore others intention or hide their real goal. In particular, informed cooperators ($I_C$) are capable to recognize correctly a partner's intention and cooperate with other cooperators while avoid being exploited by all types of defectors. Consequently, its appropriate defector version, called as informed defector ($I_D$), can only deceive unconditional cooperators. These two strategies make a special effort for additional information, which fact is considered via an extra cost. The corresponding payoff matrix is

\vspace{0.2cm}

\begin{tabular}{r|c c c c}
 & $C$ & $D$ & $I_C$ & $I_D$\\
\hline
$C$ & $R$ & $S$ & $R$ & $S$\\
$D$ & $T$ & $P$ & $0$ & $P$\\
$I_C$ & \,\,\, $R-\epsilon$ \,\,\,& $-\epsilon$ \,\,\,& $R-\epsilon$ \,\,\,& $-\epsilon$\\
$I_D$ & \,\,\, $T-\epsilon$ \,\,\,& $-\epsilon$ \,\,\,& \,\,\, $-\epsilon$ \,\,\,& $-\epsilon$\\
\end{tabular}

\vspace{0.2cm}

Here $\epsilon$ denotes the mentioned extra cost for additional information while the rest of parameters are identical those of the traditional prisoner's dilemma game. One may claim some conceptual parallel with the model discussed in previous case study, but here we have a significantly richer system behavior. More precisely, two independent solutions may emerge which are characterized by a closed loop of strategy rank. For example, a solution with the effective invasion $I_C \to D \to C \to I_C$ and an alternative triplet with the directed loop of $I_C \to I_D \to C \to I_C$ can also be detected. Notably, these triplets behave as a defensive alliance against an external strategy. For example, if a player of $I_C$ attacks a pure cooperator or defector player of ($I_C+D+C$) triplet then $I_C$ is capable to invade the external aggressor. 

Interestingly, the outcome of the battle of these triplets depends on how intensive the invasion flow within a certain triplet. For example, if the strategy exchange within the ($I_C+D+C$) loop is faster than the average invasion rate within the ($I_C+I_D+C$) loop than the former alliance prevails \cite{szolnoki_epl15}. This is a general phenomenon of models where strategy alliances compete in a multi-state system \cite{szabo_jpa05,perc_pre07b,szabo_pre08b,szolnoki_njp16}. 

Next we explore the possible consequence of a strategy-dependent learning activity where the referred strategy may be member of two competing loops. To allow the proper comparison with previous results we follow earlier parametrization, namely $T=1+r, S=-r, R=1$, while $P=-p$. In this way parameter $r$ denotes the strength of the dilemma and the measure of $p$ allows different alliances to dominate the system. Furthermore the applied additional cost of informed strategies was fixed at $\epsilon=0.3$. During the simulations we have used lattice size up to $4000 \times 4000$ to obtain the reliable solution that is valid in the large-size limit. The applied typical relaxation time was between $10^5 - 10^6$ $MCS$s.
 
\begin{figure*}
 \centering
 \includegraphics[width=3.5cm]{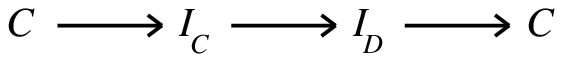}\\
 \includegraphics[width=5.5cm]{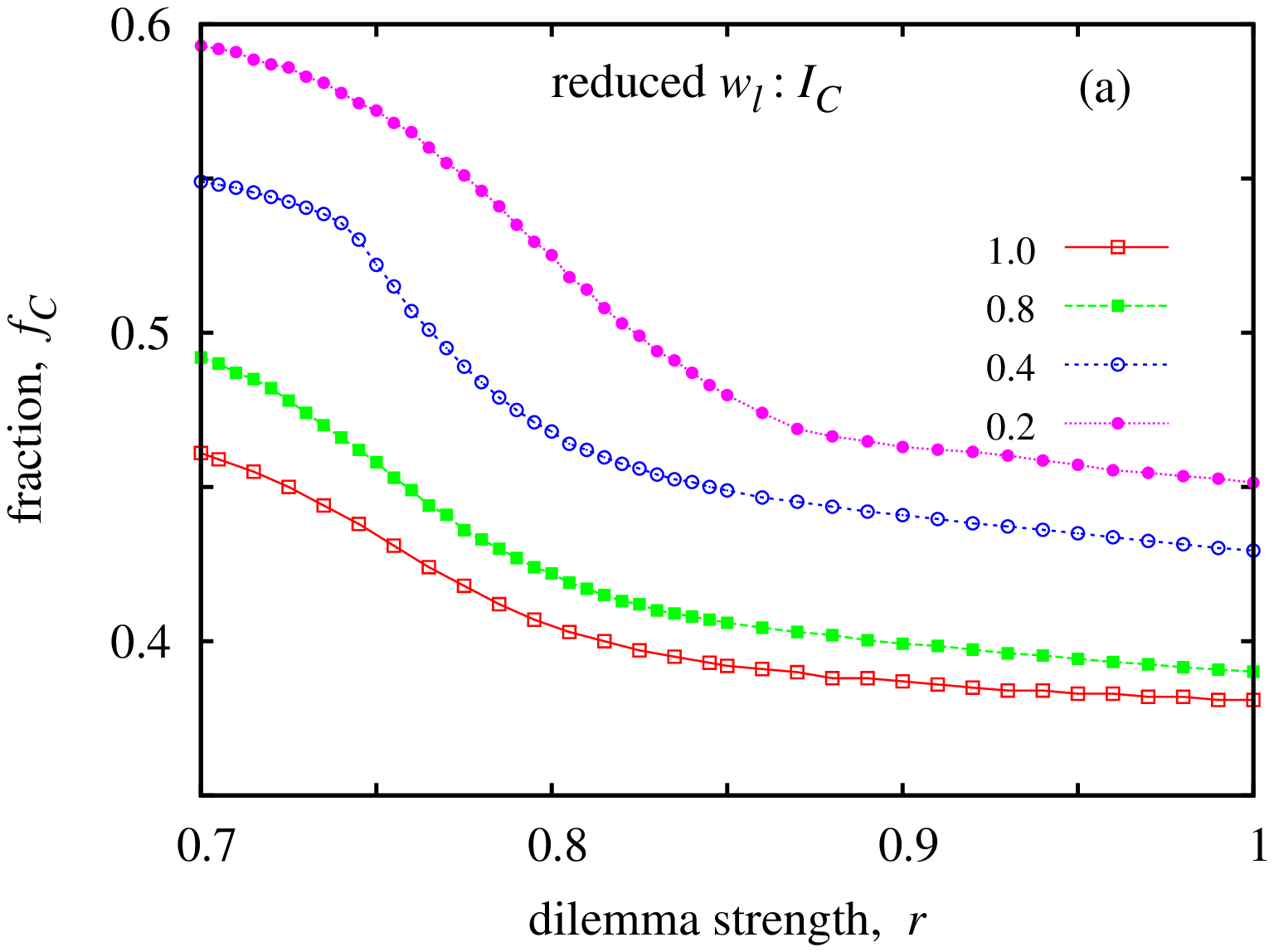}\includegraphics[width=5.5cm]{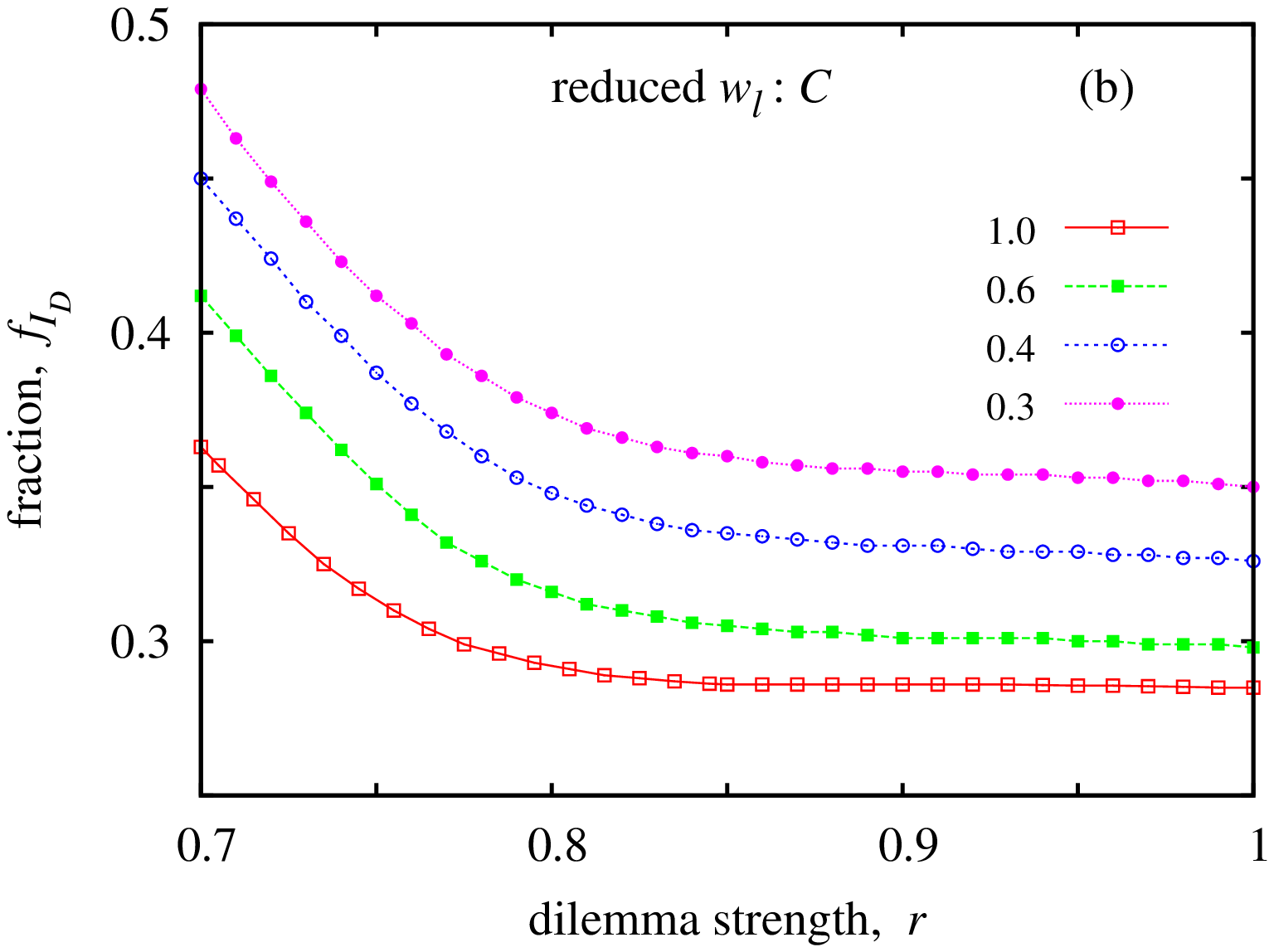}\includegraphics[width=5.5cm]{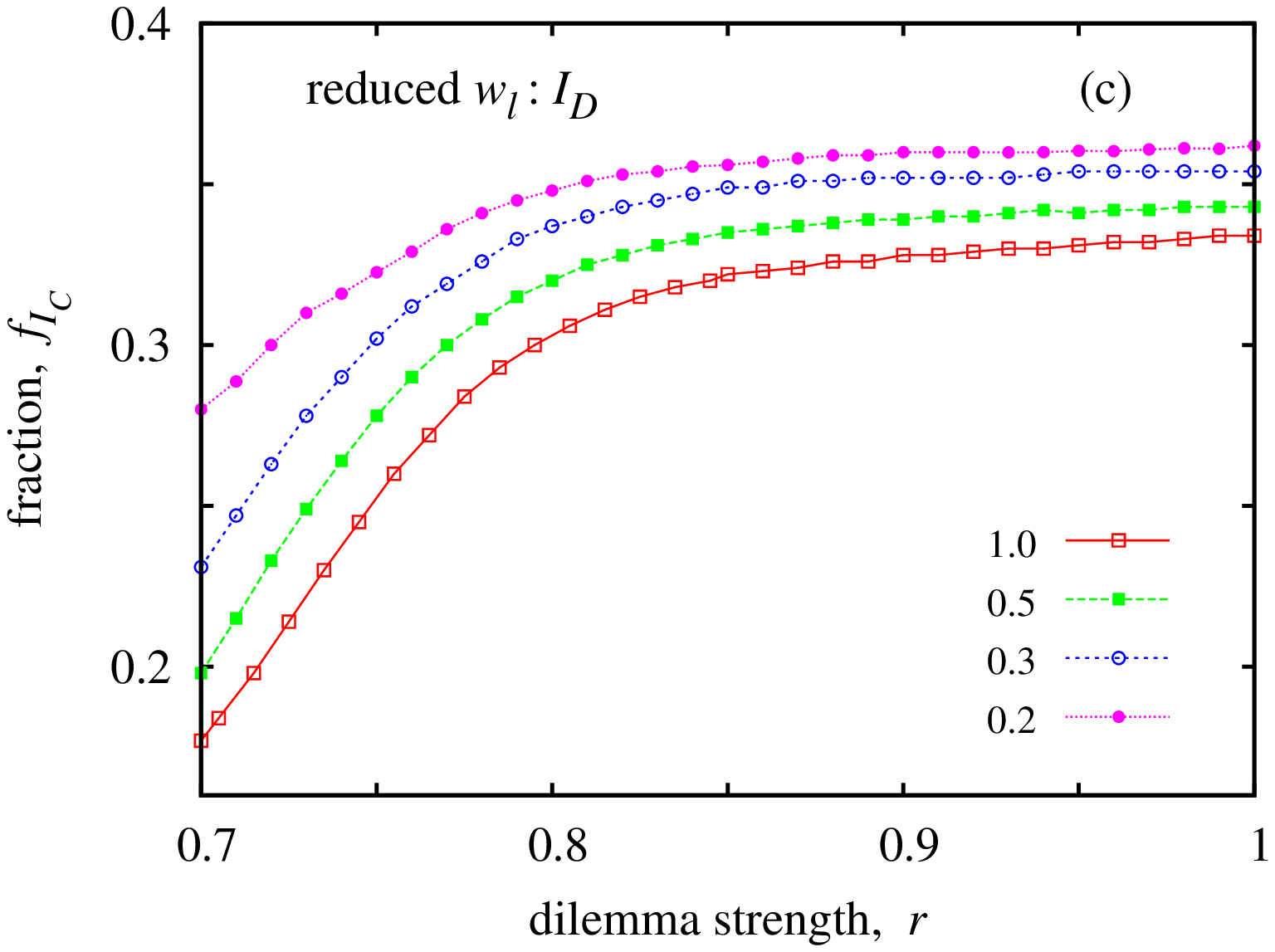}\\
 \caption{Improvement of ``predator" strategy as a function of dilemma strength in the model where both pure and informed strategies are present. Here the value of punishment is $P=-0.7$ which allows another cyclically dominant solution to emerge. The involved strategies and their ranks are indicated on the top of the plot. As usual, the applied $w_l$ values for the specific strategy are denoted in the legend. The additional cost of informed cooperators and informed defectors is $\epsilon=0.3$ for all cases.}\label{inf_II}
\end{figure*}

We first present results obtained in the light punishment ($p=0.3$) region where normally ($D+C+I_C$) alliance dominate. The results, summarized in Fig.~\ref{inf_I}, are in excellent agreement with the behavior we have observed for simpler systems previously. Namely, in every cases when the learning activity of a specific strategy is modified then the fraction of related ``predator" strategy growths independently of the strength of dilemma. In particular strategy $C$ benefits from reduced $w_l$ of $I_C$, shown in panel~(a). In the next panel $D$ enjoys the lowered learning activity of strategy $C$, and finally $I_C$ gains from the low $w_l$ values of $D$ players. Evidently the measure of enhanced frequency of the specific ``predator" strategy depends sensitively on the strength of dilemma ($r$) and the actual pair of ``predator-prey" interaction. For example, we have to decrease $w_l$ to low values to present visible improvement of $I_C$, while to increase $f_D$ was already feasible even at much significantly higher $w_l$ values of strategy $D$. It is because the effective invasion rate between strategies depends not only the learning activity, but also on the highly non-linear payoff difference of competing partners.

To close our last case study we present results obtained at harsh punishment value where the other triplet dominates in the original model. Based on our previous experience with simpler systems and the low-punishment version of present model, the results summarized in Fig.~\ref{inf_II} are in comforting agreement with our expectation. Namely, the argument that is built on ``predator--prey" relation of strategy pairs originated from cyclic dominance is so strong ingredient of the model that overcomes any other details of dynamical rule change. That is even if the intervention onto the learning activity ensures a symmetrical change toward both directions, but the system reaction recalls those behavior that was previously observed for uni-directional invasion rule. In this way, as shown in panel~(a), pure $C$ benefits from modified $I_C$ learning activity, while pure $D$ enjoys reduced $C$ learning. Last, as illustrated in panel~(c), the success of $I_C$ requires a modified $D$'s learning rule. 
 
\section{Conclusions}

We would like to stress that non-transitive relations among competing strategies are more widespread than one may na{\"\i}vely expected and they are not restricted to the Lotka-Volterra-, or $RSP$-type systems \cite{intoy_jsm13,szolnoki_pre14c,knebel_prl13,brown_pre19,nagatani_pa19b,szolnoki_pa18,baker_jtb20}, but can be potentially found in broader range of evolutionary game models \cite{cheng_f_pa19,yang_hx_pa19,fu_mj_pa19,zhang_lm_epl19,xu_zj_c19,yang_gl_pa19,liu_rr_amc19}. In general, when we increase the number of available strategies the mentioned relation among competitors emerges almost inevitably \cite{bazeia_epl18,park_c19b,esmaeili_pre18}. This is why it is crucial to clarify whether it is possible to transfer our knowledge about cyclic-dominant systems to broader models where the strategy invasion happens in both directions.

In our present work we have explored how the cyclic dominance that emerges as an effective relation between strategies influence the stationary fractions when invasion rates are modified via the microscopic adoption rule. Interestingly we have done it in a symmetric manner because the original adoption probabilities of other external strategies are reduced uniformly. In this way not just the ``predator--prey" transition is weakened, but also the learning rate from alternative strategies, too.

Despite the obvious difference from the dynamical rule of Lotka-Volterra-type systems the intervention into learning rates of strategies of a social dilemma result in conceptually similar system reaction we have already observed for the former system. We have illustrated this behavior for different models starting from the simplest, where cooperators and defectors compete with loner, to more complex systems where more than three strategies are available and they are more sophisticated than unconditional strategies.

Our observations are clear and show toward the same direction. Namely, the presence of cyclic dominance among competitors is a very strong condition that dominates the system behavior no matter invasions may happen to both directions in the loop. Consequently, when we want to reinforce a strategy by making harder to change its state then we support its ``predator" strategy unintentionally. The more we suppress a specific learning activity of an agent or species the better will benefit from it by an alternative partner. This fact should be considered in every ecological or human systems when our goal is to modify the portions of system members by an intervention into the microscopic process \cite{rafikov_ec14,roman_jtb16,park_c18c,ruifrok_tpb15}.

\vspace{0.5cm}

This research was supported by the Hungarian National Research Fund (Grant K-120785) and by the National Natural Science Foundation of China (Grants No. 61976048 and No. 61503062).


\bibliographystyle{elsarticle-num-names}

\end{document}